\documentclass[namedreferences]{solarphysics}
\usepackage[hyperref,optionalrh,showbiblabels]{spr-sola-addons} 
\usepackage{graphicx}        
\usepackage{color}           

\chardef\us=`\_

\usepackage{xcolor}
\usepackage[normalem]{ulem}
\begin{document}

\begin{article}
\begin{opening}

\title{Three Dimensional Simulations of Solar Wind Preconditioning and the 23 July 2012 Interplanetary Coronal Mass Ejection\\ {\it Solar Physics}}

\author[addressref={aff1},corref,email={ravindra.desai@imperial.ac.uk}]{\inits{R. T.}\fnm{Ravindra T.}~\lnm{Desai}\orcid{0000-0002-2015-4053}}


\author[addressref={aff2}]{\inits{H.}\fnm{Han}~\lnm{Zhang}\orcid{0000-0002-7651-1179}}

\author[addressref={aff1}]{\inits{E. E.}\fnm{Emma E.}~\lnm{Davies}\orcid{0000-0001-9992-8471}}
\author[addressref=aff1]{\inits{J. E.}\fnm{Julia E.}~\lnm{Stawarz}\orcid{0000-0002-5702-5802}}
\author[addressref={aff3}]{\inits{J.}\fnm{Joan}~\lnm{Mico-Gomez}\orcid{0000-0003-3901-3506}}
\author[addressref={aff4}]{\inits{P.}\fnm{Pilar}~\lnm{Iv\'a$\mathbf{\tilde{n}}$ez-Ballesteros}\orcid{0000-0002-7358-0486}}

\address[id=aff1]{Blackett Laboratory, Imperial College London, London, UK}
\address[id=aff2]{Department of Physics, University of Durham, Durham, UK}
\address[id=aff3]{Universitat de Val\'encia, Val\'encia, Spain}
\address[id=aff4]{Universidad Aut\'onoma de Madrid, Madrid, Spain}

\runningauthor{Desai et al.}
\runningtitle{Solar Wind Preconditioning}

\begin{abstract}
Predicting the large-scale eruptions from the solar corona and their propagation through interplanetary space remains an outstanding challenge in solar- and helio-physics research. In this article, we describe three dimensional magnetohydrodynamic simulations of the inner heliosphere leading up to and including the extreme interplanetary coronal mass ejection (ICME) of 23 July 2012, developed using the code PLUTO. The simulations are driven using the output of coronal models for Carrington rotations 2125 and 2126 and, given the uncertainties in the initial conditions, are able to reproduce an event of comparable magnitude to the 23 July ICME, with similar velocity and density profiles at 1 au.  
The launch-time of this event is then varied with regards to an initial 19 July ICME and the effects of solar wind preconditioning are found to be significant for an event of this magnitude and to decrease over a time-window consistent with the ballistic refilling of the depleted heliospheric sector. These results indicate that the 23 July ICME was mostly unaffected by events prior, but would have travelled even faster had it erupted closer in time to the 19 July event where it would have experienced even lower drag forces. We discuss this systematic study of solar wind preconditioning in the context of space weather forecasting. 
\end{abstract}

\end{opening}


\section{Introduction}
     \label{S-Introduction}
\hspace{2mm} Coronal mass ejections (CMEs) are characterised by the large-scale eruption of plasma from the solar corona and release copious amounts of energy over incredibly short time periods. Upon reaching interplanetary space, CMEs, hereafter referred to as interplanetary CMEs (ICMEs), can propagate at velocities significantly greater than the ambient solar wind and will therefore be decelerated due to drag forces \citep{Cargill96}. In recent years, the phenomenon of an initial ICME clearing the path for a successive eruption has been implicated as a significant causal mechanism for producing some of the most severe space weather events on record \citep{Liu19}. There is therefore a current need to understand how solar wind preconditioning can affect an observed CME which poses a threat to operational spacecraft and ground-based infrastructure, if Earth-directed. \citep{SWPS15}.

The most extreme ICMEs are believed to occur along a power law distribution \citep{Riley12} and can possess velocities $>$2000 km s$^{-1}$, the most famous of which is the Carrington Event of September 1859 \citep{Carrington1859}. This ICME propagated Sun-to-Earth in just 17.6 hours and induced the largest geomagnetic storm on record \citep{Tsurutani03}. On 23 July 2012, however, the Solar Terrestrial Relations Observatory-Ahead (STEREO-A) spacecraft, orbiting at 0.96 au, experienced a non-Earth directed ICME \citep{Dryer12} of a magnitude comparable to the 1859 Carrington event \citep{Baker13} which arrived 18.6 hours after its eruption from the solar corona with a velocity of $\approx$2250 km s$^{-1}$ \citep{Russell13}.    
    
Two distinct reasons have been put forth to explain the rapid transit time and extreme characteristics of the 23 July 2012 event. The first reason is the CME was produced by two separate eruptions spaced 10-15 minutes apart which appeared to merge within the solar corona \citep{Liu14}. This idea was further supported by the presence of two flux ropes within the in-situ magnetic field measurements at STEREO-A where the ICME magnetic flux consequently reached a record 109 nT  \citep{Russell13}. The second reason is that the solar wind conditions ahead of the ICME had been significantly affected by events prior to 23 July \citep{Wu17}, the most significant of which \citet{Liu14} identify as a CME which erupted from the same active region at 05:30 UT on 19 July. This propagated along a similar path and is thought to have removed existing solar wind plasma and non-radially directed magnetic fields which resulted in minimal slowdown for the larger 
subsequent eruption \citep{Temmer15}. 
It has however been highlighted that large ICMEs are significantly less affected by the ambient conditions \citep{Wu05,Liou14,Cash15} which leads to the question as to if, and how, the 23 July 2012 ICME was, or could have been, influenced by prior preconditioning events?
    
The complex interaction between ICMEs and the solar wind have been analysed within a number of  simulation studies \citep[e.g.][]{Cargill96,Odstrcil99,Riley03,Lugaz05}, and \citet{Manchester08} and \citet{Werner19} include the effects of solar wind preconditioning when simulating the solar storms of 28 October 2003 and 6-9 September 2017, respectively. This study focuses on solar wind preconditioning with respect to the Carrington-scale event observed on 23 July 2012.

This article is organised as follows: Section \ref{PLUTO} describes the simulation code, PLUTO, the governing equations and implementation. Section \ref{23july2012} builds upon these concepts and describes a simulation of the conditions leading up to and including the extreme ICME of 23 July 2012. Section \ref{preconditioning} then uses this simulation as a baseline to study solar wind preconditioning for events of this magnitude. Section \ref{summary} then concludes with a discussion on the implications for understanding and forecasting these extreme events.

\section{Simulation Development}
\label{PLUTO} 
\subsection{Governing Equations}

\hspace{0.3cm}  The simulation code, PLUTO, is an open-source parallel modular multiphysics code developed for the study of astrophysical plasmas \citep{Mignone07}. It is written to solve partial differential equations deriving from conservation laws of the form,
\begin{equation}
    \frac{\partial \mathbf{U}}{\partial t} = -\nabla \cdot \mathbf{T(U) + S(U)},
\end{equation}
where \textbf{U} is the state vector of conservative quantities, \textbf{T} is a tensor denoting the flux of
each component of the state vector and \textbf{S} defines the source terms. This general integration sequence does not require the explicit forms of \textbf{U}, \textbf{T} and \textbf{S}, and this flexibility allows the implementation of different physical equations.  PLUTO has the capability to solve non-relativistic or relativistic hydrodynamic (HD) or magnetohydrodynamic (MHD) systems of equations on either cartesian, cylindrical or spherical geometries. 

PLUTO has been validated against a number of test-problems \citep[e.g.][]{Mignone07} and extensively applied to the study of astrophysical plasmas, including; stellar and extragalactic jets \citep[e.g.][]{Bodo03,Mignone13}, the solar corona \citep[e.g.][]{Reale16,Petralia18}, stellar winds \citep[e.g.][]{Alvarado16,Pantolmos17}, and 1D simulations of the fast-forward shock associated with the 23 July 2012 event \citep{Riley16}.

A detailed description of the numerics and physics capabilities are described by \citet{Mignone07} and for completeness the system of equations solved in this study are described below. 
The MHD conservation equations are written as:
\begin{equation}
    \hspace{-10mm} \frac{\partial \rho}{\partial t} + \nabla \cdot \rho \mathbf{v} = 0,
\end{equation}
\begin{equation}
    \frac{\partial \mathbf{m}}{\partial t} + \nabla \cdot \left[ \mathbf{mv} - \mathbf{BB} + \mathbf{I}(p + \mathbf{B}^2) \right]^T = -\rho \nabla \mathbf{\Phi} + \rho \mathbf{g},
\end{equation}
\begin{equation}
    \frac{\partial \mathbf{B}}{\partial t} + \nabla \times (\mathbf{v}  \times \mathbf{B}) = 0,
\end{equation}

\begin{equation}
    \frac{\partial (E_t + \rho \mathbf{\Phi})}{\partial t} + \nabla \cdot \left[ (\frac{\rho \mathbf{v}^2}{2} + \rho e + p + p \mathbf{\Phi})\mathbf{v} + (\mathbf{v}  \times \mathbf{B}) \times \mathbf{B} \right] = \mathbf{m} \cdot \mathbf{g},
\end{equation}
where $\mathbf{v}$ is the gas velocity in an inertial reference frame, $\rho$ is the gas mass density, and $p$ its thermal pressure. $\mathbf{m}$ represents the momentum density term equal to $\rho \mathbf{v}$, $\mathbf{I}$ is the unit tensor, and $\mathbf{\Phi}$ and $\mathbf{g}$ are the potential and vector part of the body force which are set to zero in this study. The total number density $n$ is defined by $\rho = \mu H$, where $\mu$ is the mean molecular weight in units of the mass of the hydrogen atom $H$. $\mathbf{B}$  
is the magnetic field and the total energy density, $E_t$, is defined as,
\begin{equation}
    E_t = \rho e + \frac{\mathbf{m}^2}{2\rho} + \frac{\mathbf{B}^2}{2}.
\end{equation}
In Equations 4 and 5 the electric field $\mathbf{E}$ is provided through the generalised Ohm's law with the resistive and Hall terms neglected, leaving an ideal MHD description. A closure is provided to these systems using an equation of state $\rho \hspace{1mm} e = \rho \hspace{1mm} e (p,\rho )$, which is set to represent an ideal gas. 

To evolve these equations in time, the code uses volume averages determined by piece-wise monotonic interpolation inside each grid cell. 
This generic implementation allows the code to easily switch between a number of different solvers and, in this study, the Roe solver of
\citet{Cargo97} is utilised. The Linearized Roe Riemann solver is based on characteristic decomposition of the Roe matrix which approximates the solutions to the discontinuous left and right Riemann solutions at the cell interfaces \citep{Roe81}. The MHD formulation evolves $\rho$, $\mathbf{v}$, $T$ and $\mathbf{B}$ in time and  requires an additional step to enforce the  solenoidal constraint on the divergence of the magnetic field. The eight-wave divergence cleaning formulation of \citet{Powell94} and \citep{Powell99} is used, with the constrained transport scheme of \citet{Balsara99} and \citet{Londrillo04}, to add an additional corrective source term. 
This utilises the upwinded fluxes in the higher-order Godunov scheme combined with the electric field to maintain a divergence-free magnetic field to machine precision.

\hspace{0.3cm} The heliospheric simulations can be implemented in one, two or three spatial dimensions with either one, two or three components of the vector fields. 
In this study, a spherical coordinate system is used where $r$, $\theta$ and $\phi$ represent the radial, polar and azimuthal directions respectively, in a heliocentric rotating or inertial reference frame. The mean molecular weight is set to $\mu$=0.6 to represent a fully ionised electron-proton plasma with small amounts of heavier species. The domain is specified to capture the solar wind expanding outward from an inner boundary specified at an arbitrary distance from the solar surface. The simulations are driven by updating this inner inflow boundary at each time-step with commensurate solar wind values. For idealised scenarios the inputs are theoretically derived, or, to represent real time periods, the values at the inner boundary are updated to match the output of coronal models.
The outer boundaries are represented by Von-Neumann boundary conditions for continuous outflow, and the simulation domain thus represents the continuous evolution of the solar wind within a predefined region of the heliosphere.

  \begin{figure}    
   \centerline{\includegraphics[width=1.4\textwidth,clip=]{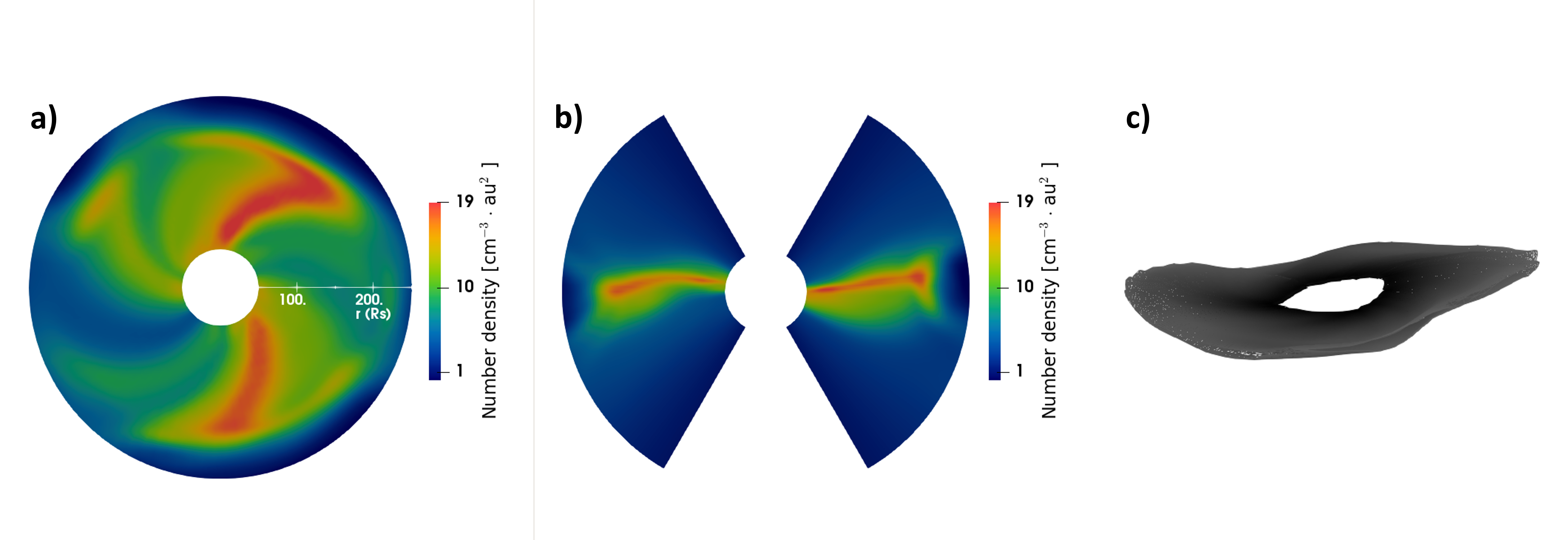}
              }
              \caption{Initialisation phase of a three dimensional MHD simulation of the solar wind. The simulations are shown as the fast solar wind first reaches the outer boundary while the slow solar wind has not yet done so. The left-hand and centre panels (a--b) show the plasma number density normalised by the radial distance squared in au from above and at the equatorial plane, respectively. The right-hand panel (c) shows the heliospheric current sheet which is revealed by masking magnetic field values above an arbitrarily small value. Note that 214.8 R$_s$ $\simeq$ 1 au }
   \label{solarwind}
   \end{figure}

\subsection{The Solar Wind}
\label{setup}

\hspace{0.3cm}  To simulate the nascent solar wind, a three-dimensional MHD simulation is initialised to reproduce solar wind conditions starting from 21 June 2012. The simulations are driven using inflow coronal boundary conditions generated by CORHEL \citep{Riley01,Linker99} from the archived simulation run \citet{Scolini12}, downloaded from NASA’s Community Coordinated Modeling Center (CCMC). 
This CORHEL coronal boundary condition uses the magnetic field structure from the Magnetohydrodynamics Algorithm outside a Sphere (MAS) polytropic coronal model \citep{Linker99} with empirically derived velocity, density and temperature variables \citep{Riley01}. The input data is available at a 6 hour cadence and a resolution of $\theta=180$ and $\Phi=360$ points. A cubic spline interpolation sequence is used to resolve this to the variable simulation timestep of the heliospheric grid which is specified at a lower resolution of $\theta=90$ and $\Phi=180$ points with a radial extent of 192 points from 30 Rs out to beyond 1 au.

Figure \ref{solarwind} shows the initialisation stage of this simulation where the solar wind expands outward into a low density background. The plasma density across the simulation domain is initially specified to fall off as a function of $r^{-2}$, but the simulations are only judged to have reached a steady-state when solar wind plasma from the inner boundary has reached the outer boundary and asymptoted to a steady state solution. The left and centre panels show two projections of the normalised density where the fast solar wind has reached the outer boundary but the slow solar wind has not yet done so. The propagation of fast and slow solar wind and stream interaction regions (SIRs) are visible as well as the rotation of the Sun which wraps these streams along an Archmidean spiral \citep{Parker58}.
The right-hand panel shows the heliospheric current sheet (HCS), calculated by masking values above an arbitrarily small value. The HCS is produced at the Sun's magnetic equator and maintains its tilted orientation with respect to the ecliptic plane as it evolves outward. 

In this study a polytropic index of  $\gamma = 1.5$ is used to capture the measured polytropic index of the free-streaming solar wind \citep{Totten95}. This still, however, doesn't fully capture the kinetic heating of the solar wind via a variety of electron and ion-scale instabilities \citep[e.g.][]{Stawarz09,Goldstein15} and the MHD simulations therefore result in a steeper radial temperature profile than that observed \citep[e.g.][]{Cranmer09}.

\subsection{Interplanetary Coronal Mass Ejections}
\label{ICME}

\hspace{3mm} CMEs are launched along a specific direction and are then able to deflect and rotate as they erupt outward \citep{Gopalswamy03, Cremades06, Gui11,Vourlidas2011} as they interact with the solar wind \citep{Wang04,Isavnin13} or through interactions with other ICMEs \citep{Lugaz12,Shen12}. 
This can cause limb ICMEs to strike Earth \citep{Wang04} and 30-40\% of Halo ICMEs to not strike Earth \citep{Yermolaev05,Shen14}.

To model and capture these three dimensional processes, a pulse is injected on the inner boundary by varying the solar wind density and velocity according to the relation:
\begin{equation}
    \lambda(t,\theta,\phi) = \lambda_0 + \lambda_p sin^2(\pi \frac{t - t_0}{t_p}) cos(\frac{\pi}{2}\frac{\theta - \theta_0}{\Theta}) cos(\frac{\pi}{2}\frac{\phi - \phi_0}{\Phi}) ,
\end{equation}
within the limits:
\begin{equation}
    t_0 < t_p < t_0 + T, \hspace{5mm} \theta_0 - \Theta < \theta < \theta_0 + \Theta,  \hspace{5mm} \phi_0 - \Phi < \phi < \phi_0 + \Phi,
\end{equation}
where $\lambda$ represents the variable of interest, and $\lambda_p$ its peak increase. $\theta_0$, $\phi_0$ and $t_0$ are the latitudinal and azimuthal launch position and start time, respectively, and $\Phi = \Theta$ which represents the half-angle subtended by the pulse and $t_p$ the pulse duration. 

Gaussian and sinusoidal pulses, and variations thereof, have been commonly used to represent ICMEs \citep[][]{Odstrcil99,Pommoell18,Liou14,Riley16} and, as these pulses propagate outward, the outer regions expand faster than the inner regions and produce a characteristic cone shape with associated MHD forward and reverse shocks. These pulses do not strictly conserve the Rankine-Huguenot jump conditions but are of sufficient smoothness that MHD solvers are able to relax these into a structure which has been shown to well-represent an ICME \citep[e.g.][]{Chane06}, with the exception of the internal flux rope magnetic field.

  \begin{figure}    
   \centerline{\includegraphics[width=1.2\textwidth,clip=]{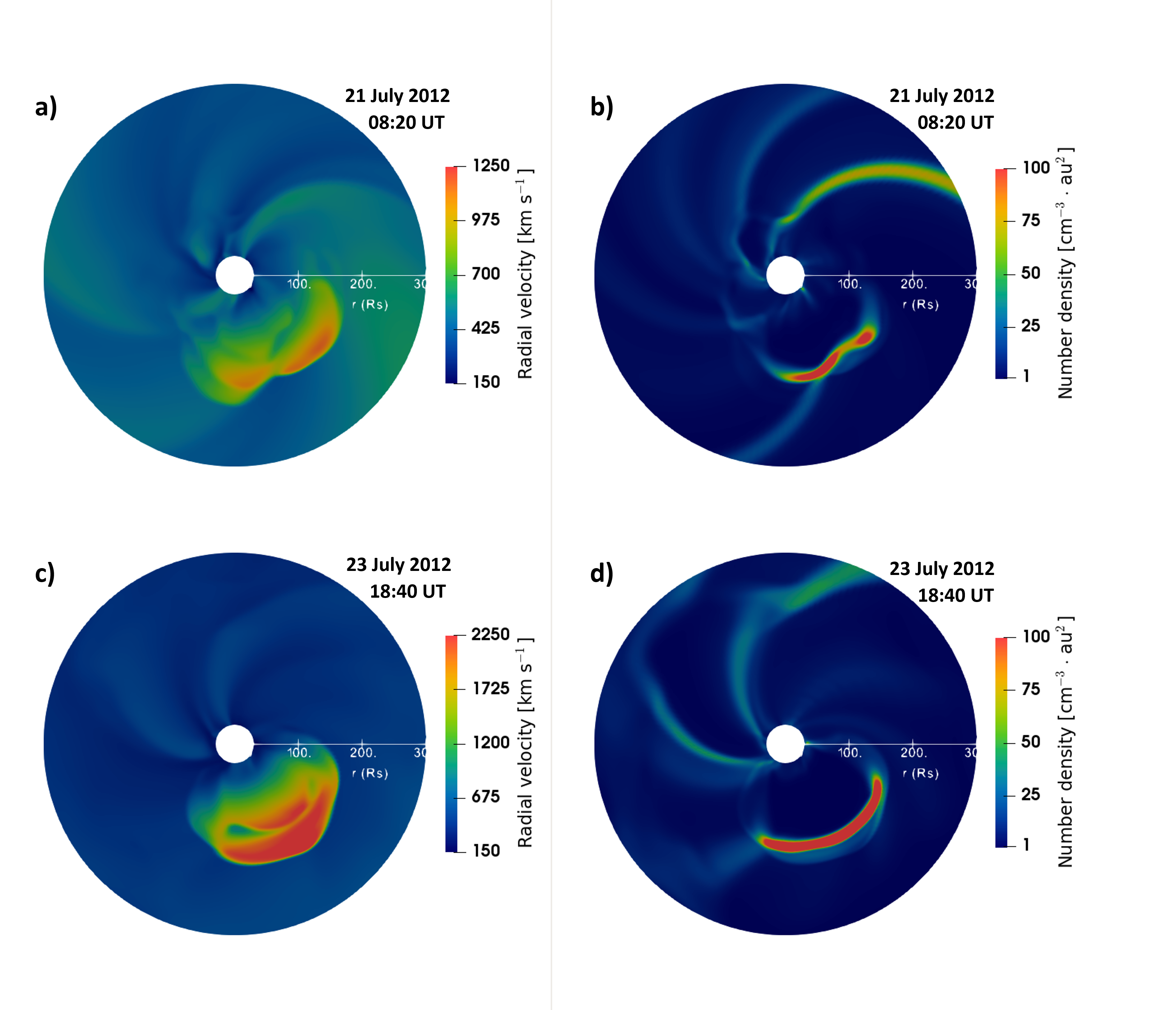}8
              }
              \caption{Three dimensional MHD simulations of the 19 July ICME in upper panels (a--b) and 23 July ICME in lower panels (c--d). The left-hand  panels (a and c) show the radial velocity and the right-hand panels (b and d) show the scaled plasma number density. Note that 214.8 R$_s$ $\simeq$ 1 au. }
   \label{mainevent}
   \end{figure}

\section{Simulating the 23 July 2012 Event} \label{23july2012}

\hspace{5mm} In this study the preconditioning events leading up to 23 July and the 23 July ICME are approximated by two distinct CMEs erupting on 19 July and 23 July. The background solar wind is modelled for the time period encompassing this event from 15 July, thus allowing the simulated heliosphere to reach steady-state before the passage of the ICMEs. The simulations are driven using two publically available simulations \citep{Mannucci12, Scolini12} produced by the MAS polytropic coronal model \citep{Linker16} which cover Carrington rotation 2125 from 21 June until 18 July, and Carrington rotation 2126 from 18 July until 16 August. These coronal simulations are driven by GONG synotoptic magnetogram data. To avoid numerical instability, the final outputs of Carrington rotation 2125 are omitted and the model inputs are interpolated over 24 hours, which ensures a smooth transition between changes in the locations of the coronal hole boundary between the two inputs.

  \begin{figure}    
   \centerline{\includegraphics[width=1.1\textwidth,clip=]{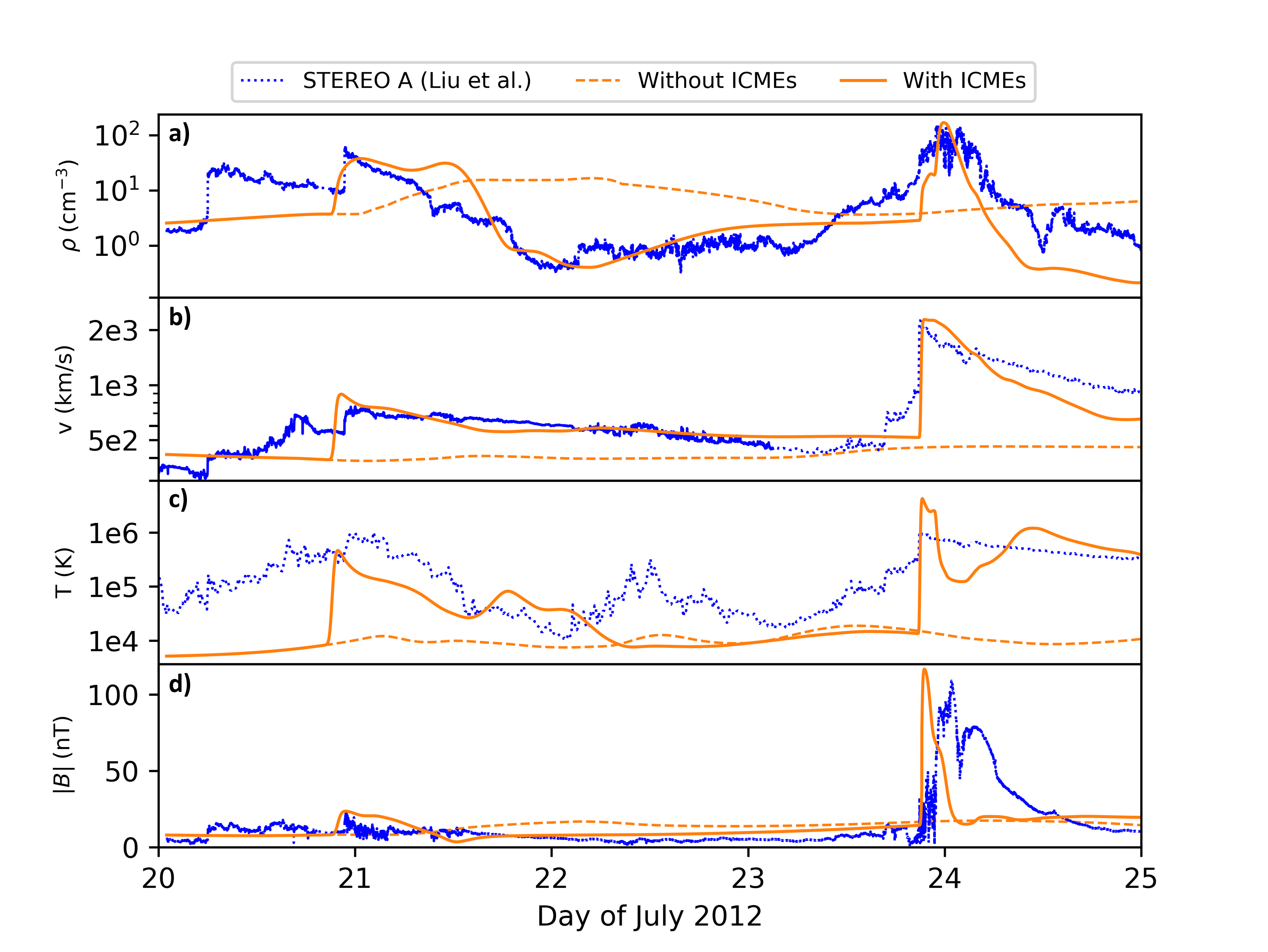}
              }
              \caption{Time series output from the MHD simulations presented in Figure \ref{mainevent} at the location of STEREO-A compared to the STEREO-A timeseries from \citet[][see Figure 5 therein]{Liu14}.  (a) Plasma density, (b) bulk velocity, (c) temperature and (d) magnitude of the magnetic field. Note, the total measured density (a) (blue) were calculated through extrapolation from the low energy electron measurements \citep{Liu14}.  }
   \label{stereoA}
   \end{figure}

The ICMEs are initialised in-line with measurements \citep{Cash15,Liou14}, see Table 1, and the simulations run until the final eruption has reached the outer boundary on 24 July 2012. 
Estimates used for the initial speed of the 23 July ICME range from 2500$\pm$500 km s$^{-1}$ to $>$3000 km s$^{-1}$ \citep{Baker13,Liu14,Intriligator15,Liou14,Intriligator15,Riley16,Liu17}, the upper range of which exceeds the theorised $\approx$ 3000 km/s limit available from the energy budget of solar active regions \citep{Gopalswamy05,Yashiro04}. In this study a launch velocity of 2800 km/s is found to produce a reasonable match.

Figure \ref{mainevent} shows the simulation of the solar wind and 19 July and 23 July ICMEs. The 19 July ICME is visibly smaller than the 23 July ICME and the initial eruption produces a density cavity in its wake. The ICMEs propagate through various solar wind conditions; the 19 July ICME encountering a denser slower solar wind stream, which cause significant azimuthal variation across solar longitude, whilst the 23 July ICME propagates from a fast solar wind stream into a slow solar wind stream. The limitations in using two distinct coronal models inputs become clear as in certain locations, the changes in the coronal hole boundaries between the two inputs causes fast solar wind to run into slower solar wind which results in some visible discontinuities. 
Further interpolation routines might remove this and indeed \citet{Cash15} implement a variety of different solar wind backgrounds from either before or after the event and finally a hydrodynamic solar wind. It is not necessarily clear, however, which selection is correct due to the lack of continuous magnetogram coverage during this period. These effects are also minimal in the sector where the 19 and 23 July ICMEs propagate, and the ICME propagation along the Sun-STEREO-A line is not significantly affected. 

Figure \ref{stereoA} shows the time series from a probe within the simulation at the location of STEREO-A which is compared to the STEREO-A in-situ measurements. The time series is also shown for a simulation of the background solar wind without the ICMEs.  Two initial ICMEs are visible in the STEREO-A data on 18 and 19 July which in this study are represented by the single and larger 19 July ICME. The 19 July ICME appears to feature a second peak in density which is attributed to the dense solar wind stream encountered en-route which is likely to have been cleared by the first initial ICME visible in the STEREO-A time series, but which is not simulated here. The low density cavity ahead of the 19 July ICME is well reproduced with the background plasma density dipping to comparable values to the observed $\approx$1 cm$^{-3}$.  The velocity profile of the 23 July ICME is subsequently a close match with the simulated ICME arriving close in time to the observed ICME, and initially dropping of at a similar rate. A second peak in velocity is visible in Figure \ref{mainevent}, which represents a reverse shock, although this feature is only marginally visible within the measurements. The density peaks at a similar time and amplitude and, with the velocity profile, indicates that the simulations have reproduced an ICME of comparable magnitude to the observed event. 

Notable differences do however exist. The initial ramp-up in density and velocity for the few hours prior to the 23 July ICME is not captured, some of which \citet{Russell13} tentatively attribute to initial disturbances from the 23 July event itself. The simulated ICME density is also narrower in time, reaching higher values but decreasing more rapidly too. 
The simulated magnetic field reaches $>$100 nT, as in the STEREO-A time-series, but this is due to compressed pre-existing solar wind fields rather than an intrinsic field and therefore decreases rapidly too. These discrepancies are attributed to the approximation of a complex multi flux rope event by a single smooth pulse. The simulated 23 July ICME also reveals a distinct temperature spike not captured within the in-situ measurements.

Further efforts to optimise the input parameters to match the observations could decrease the discrepancies but might well result in over-fitting due to the various error sources deriving from the choice in background solar wind conditions and the observational constraints on the initial and final conditions. Given these, the concept of a high-speed ICME travelling through a low density interplanetary medium is judged to be well captured and the simulations presented in Figures \ref{mainevent} and \ref{stereoA} are thus taken as a reasonable reproduction of the 23 July 2012 event, and indeed of a Carringtons-scale event, from which physical processes can be inferred.

  \begin{figure}    
   \centerline{\includegraphics[width=1.4\textwidth,clip=]{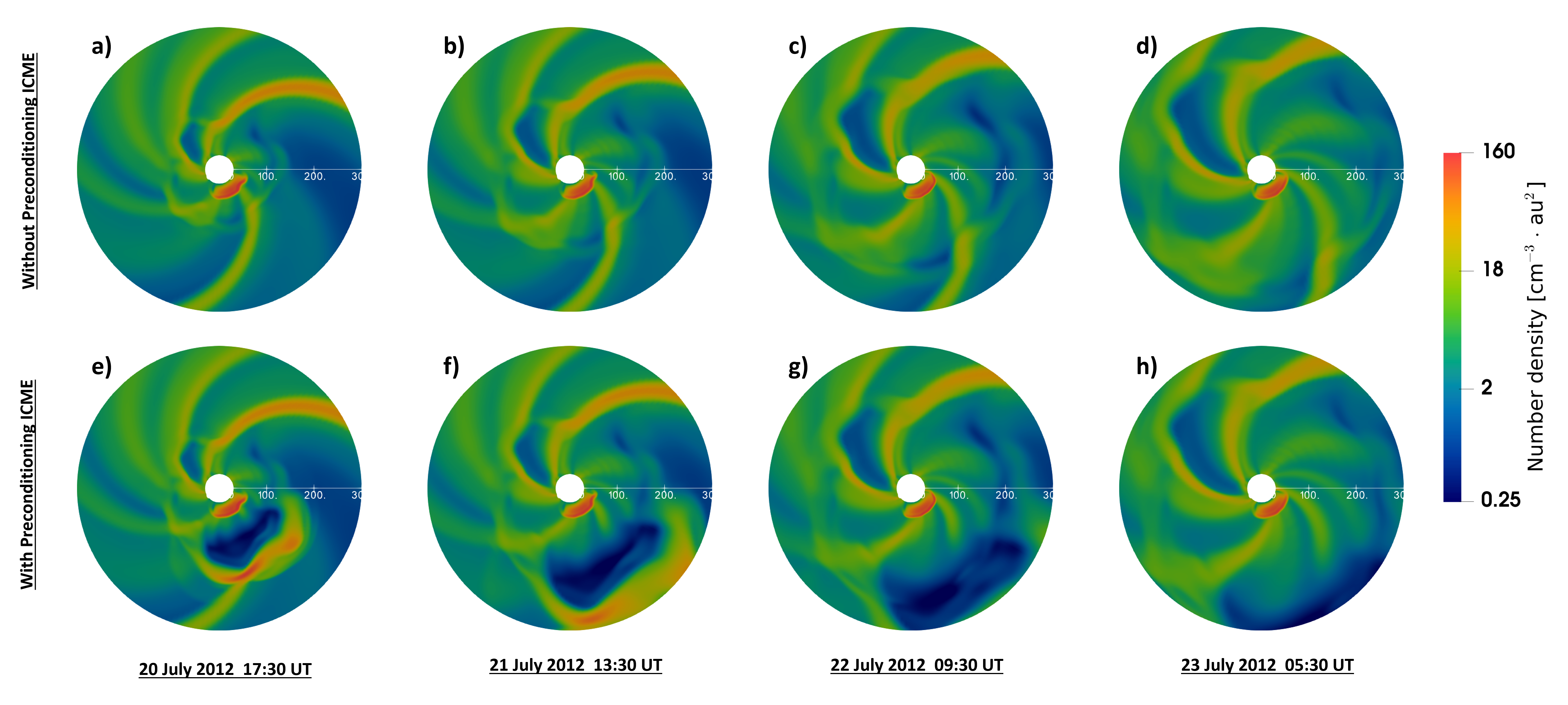}
              }
              \caption{Normalised plasma number density for four different launch times from 20\,--\,23 July. Upper panels (a--d) show the case of a single ICME without solar wind preconditioning and lower panels (e--h) show the case of a preceding ICME which produces a large density cavity in front of the subsequent event. The right-hand lower panel (h) corresponds to the simulations shown in Figures \ref{mainevent} and \ref{stereoA}. Note that 214.8 R$_s$ $\simeq$ 1 au. }
   \label{8plot}
   \end{figure}
   
\section{Preconditioning Study} \label{preconditioning}

   \begin{figure}    
   \centerline{\includegraphics[width=1.1\textwidth,clip=]{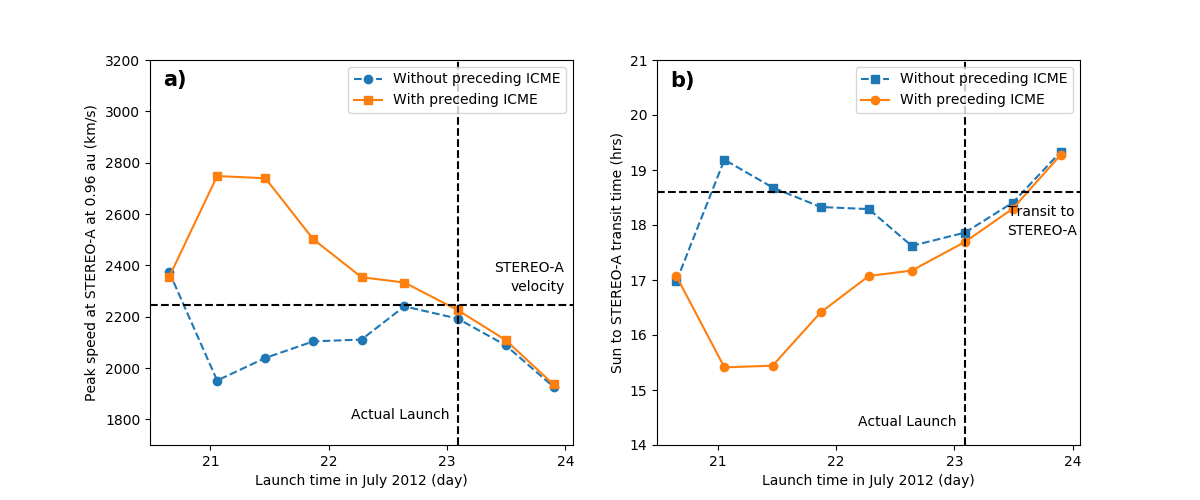}
              }
              \caption{Parametric study showing the variations in radial velocity in left-hand panel (a), and transit time in right-hand panel (b) resulting from the 23 July ICME launched at different times behind the 19 July ICME (orange solid line) and also in its absence (blue dashed line). The actual velocity observed at STEREO-A and transit time are marked by annotated dashed black lines. }
   \label{launchtime}
   \end{figure}  

\hspace{2mm} The simulated times series in Figure \ref{stereoA} show the solar wind density ahead of the main event to increase towards their nominal levels faster than the observations do although they still have not fully recovered by 23 July. This faster recovery appears to be due to the higher simulated solar wind speeds ahead of the 23 July ICME of $\approx$550 km s$^{-1}$ which gives a ballistic refilling timescale of $\approx$3.1 days for fresh solar wind to propagate into the depleted sector, following the passage of the 19 July ICME. The STEREO-A observations, however, show a lower velocity of $\approx$440 km s$^{-1}$, which gives a ballistic refilling timescale of $\approx$3.9 days which extends up until the 23 July ICME. The ballistic refilling argument does, however, only provide a minimum timescale for the refilling as it does not account for the density cavity ahead of the fresh solar wind and how the remaining rarefied solar wind has been impacted by the initial event.
\citet{Temmer17} indeed find that the interplanetary medium remains disturbed for $\approx$ 3\,--\,6 days based on statistical comparisons between observed ICMEs and modelled background solar wind.

To further study the concept of an initial ICME clearing the path for a second ICME, the simulation presented in Figures \ref{mainevent} and \ref{stereoA} are used as a baseline and the launch time of the 23 July ICME is varied with respect to the 19 July ICME, from 20 July until 24 July. The simulations are also run in the absence of an initial event to fully isolate the effects of solar wind preconditioning. From an operational forecasting perspective this approach allows one to examine the amount of time after an initially observed ICME that subsequent ICMEs will have an enhanced geophysical impact. Figure \ref{8plot} shows four of these permutations, with and without preconditioning, which reveals the significant density cavity ahead of the subsequent ICME and how an earlier launch time results in the subsequent ICME propagating through this cavity. 

The peak velocity at the location of STEREO-A is shown in Figure 5 with the simulation time corresponding to the actual launch marked. The peak velocity, together with the southward component of the magnetic field, provides an indicator of the geoeffectiveness of the event \citep{Burton75,Wu05b}, while the transit time is also shown which provides the lead time that is required to accurately predict a given CMEs evolution. The results show that the initial simulated 19 July ICME induced a speed increase in the subsequent event right up until 24 July, although, on and after 23 July this effect appears minimal as the simulations with and without the 19 July ICME converge. 
As the simulated 23 July ICME is launched earlier in time, the peak velocities increase as the 23 July ICME spends more and more time in the wake of the 19 July ICME and eventually runs into its back if launched before 21 July, a phenomenon \citep{Lugaz05} not examined in this study . The velocity increase appears the most significant when launched on 21 July.

The preconditioning study results, combined with further analysis of the STEREO-A data suggest that the preconditioning for the 23 July 2012 event was marginal and therefore this event was produced without the aid of solar wind preconditioning. The trends do, however, indicate that solar wind preconditioning can be significant for an event of this magnitude and that this event therefore could have been even more severe if launched directly into the depleted solar wind observed by STEREO-A two days prior.

\section{Discussion and Conclusions\label{sec:Conclusion-and-Future}}
\label{summary}
\hspace{5mm}

This study describes 3D MHD PLUTO simulations of the inner heliosphere leading up to and including the 19 and 23 July ICMEs, the latter of which has been identified as a Carrington-scale event. The simulations were driven by CORHEL coronal boundary conditions for Carrington rotation 2125 and 2126, which highlighted the relevance of continuously derived inputs with respect to observing and modelling the solar corona. The simulations produced an event of comparable magnitude to the 23 July 2012 event with time-series output from the location of STEREO-A matching the observations. It should, however, be noted that the simulations of the 23 July 2012 event are constrained around the STEREO-A observations at 0.96 au. It is therefore likely that multiple initial conditions exist that produce an ICME at 1 au that matches those observed, as indeed is highlighted by \citet{Cash15}. The uncertainties in the near-Sun properties of the CME do not allow for this event to be constrained further and simulations of this events should therefore be interpreted as physically realistic events of comparable magnitude to the 23 July 2012 event rather than exact replicas.

The 23 July 2012 simulation was then used to further examine and quantify the impact of solar wind preconditioning. The launch time of the 23 July ICME was varied with respect to a 19 July preconditioning ICME and also in its absence to constrain the phenomenon of one ICME clearing the path for another. The simulations were able to reproduce the observed depleted solar wind densities ahead of the 23 July ICME and the simulated 23 July ICME reached a maximum velocity of $>$2750 km s$^{-1}$ when launched on 21 July but tended towards 2000 km s$^{-1}$ when launched after 23 July.

Several studies have indicated that large ICMEs are significantly less affected by the ambient solar wind \citep{Wu05,Cash15}. The simulations of the 23 July ICME presented here are, however, towards the larger end of those considered, see Table 1 and \citet[][Table 2. therein]{Cash15}. While the precise numbers presented in this parametric preconditioning study might well differ depending on the solar wind backgrounds and ICME parameters used, the depletion behind the 19 July ICME shown in Figure \ref{8plot} is consistent with the large-scale interaction between ICMEs and the ambient solar wind and also comparable to those produced by flux-rope ICME simulations \citep[][]{Shen14}. The trends presented in Figure \ref{launchtime} therefore indicate that solar wind preconditioning can be a significant factor for ICMEs of comparable magnitude to the 23 July event and, therefore, Carrington-scale events. The simulations also reveal that the 23 July event itself likely occurred when the solar wind had nearly fully recovered and it can be inferred that the 23 July would  have reached even higher velocities at 1 au if launched closer in time to the 19 July preconditioning ICME. 

The preconditioning window identified in Figure \ref{launchtime}, although focussed on a particular event, can provide context for historical events, such as the August 1972 ICME which reached Earth in a record 14.6 hours \citep{Knipp18} and the Carrington event of September 1859 which reached Earth in 17.6 hours \citep{Tsurutani03}, which have been implicated as involving multiple ICME eruptions. In the absence of in-situ measurements, the trends presented here indicate that preconditioning variations in the ambient solar wind could explain the difference in these arrival times. The significant increase in ICME velocity and decrease in transit time due to solar wind preconditioning could also plausibly enhance the severity of a range of CMEs which do not initially present themselves as Carrington-scale within the solar corona. This study also did not, however, consider the effects of solar wind preconditioning inside the solar corona. A comprehensive parametric preconditioning study, with regards to the numerous ICME input parameters that can be varied, and also with a flux rope CME description, is therefore left for a future endeavour.

The preconditioning window induced by an initial ICME, and the decrease in its significance over time, presents a quantifiable prediction that can be used by current space weather forecasting efforts to estimate the increased likelihood of a severe space weather event occurring.

\vspace{1cm}

 
 \begin{table}[h]
\begin{tabular}{c|ccccccc}
\textbf{ICME}    & \textbf{\begin{tabular}[c]{@{}c@{}}Launch Time\\ (UT)\end{tabular}} & \textbf{\begin{tabular}[c]{@{}c@{}}Radial Velocity\\ (km s$^{-1}$)\end{tabular}} & \textbf{\begin{tabular}[c]{@{}c@{}}Mass \\ (g)\end{tabular}} & \textbf{\begin{tabular}[c]{@{}c@{}}Longitude\\ ($^{\circ}$)\end{tabular}} & \textbf{\begin{tabular}[c]{@{}c@{}}Latitude\\ ($^{\circ}$)\end{tabular}} & \textbf{\begin{tabular}[c]{@{}c@{}}Half Width\\ ($^{\circ}$)\end{tabular}} & \textbf{\begin{tabular}[c]{@{}c@{}}Duration \\ (hrs)\end{tabular}} \\ \hline
\textbf{19 July} & 05:30                                                               & 1500                                                                             & 1.0e+16                                                      & 134                                                                       & 0                                                                        & 45                                  & 3.5                                        \\
\textbf{23 July} & 04:30                                                               & 2800                                                                             & 2.0e+16                                                        & 134                                                                       & 0                                                                        & 45 
& 3.5
\end{tabular}
\caption{Input data for the 19 July and 23 July 2012 ICMEs corresponding the simulations presented in Figure \ref{mainevent} and \ref{stereoA}. The ICMEs are launched at the inner boundary of 30 R$_S$ and further input values are taken from \citet[][see Tables 1 \& 2 therein]{Cash15}.
 \label{inputtable}}
\end{table}

\begin{acks}
The authors would like to extend their gratitude to Robert Forsyth, Jeremy Chittenden, Jonathan Eastwood, Timothy Horbury, Peter Riley and Jon Linker for useful discussions, and Ying Liu for providing updated STEREO-A time-series estimates. RTD acknowledges funding from the NERC grant NE/P017347/1 (Rad-Sat).
EED acknowledges STFC studentship ST/N504336/1. JES acknowledges funding from the UKRI STFC grant ST/S000364/1. PI and JM acknowledge the Erasmus programme. The CORHEL coronal boundary conditions were downloaded from the Coordinated Community Modelling Centre. This study used the Imperial College High Performance Computing Service (doi: 10.14469/hpc/2232).

\vspace{1em}

\textbf{Disclosure of Potential Conflicts of Interest:} The authors declare that they have no conflicts of interest.
\end{acks}

\bibliographystyle{spr-mp-sola}
\bibliography{sola}  


\IfFileExists{\jobname.bbl}{} {\typeout{}
\typeout{****************************************************}
\typeout{****************************************************}
\typeout{** Please run "bibtex \jobname" to obtain} \typeout{**
the bibliography and then re-run LaTeX} \typeout{** twice to fix
the references !}
\typeout{****************************************************}
\typeout{****************************************************}
\typeout{}}

\end{article} 

\end{document}